\documentstyle[12pt,amssymbols]{article}
\def\journal{\topmargin .3in	\oddsidemargin .5in
	\headheight 0pt	\headsep 0pt
	\textwidth 5.625in 
	\textheight 8.25in 
	\marginparwidth 1.5in
	\parindent 2em
	\parskip .5ex plus .1ex		\jot = 1.5ex}
\journal

\catcode`\@=11
\def\marginnote#1{}
\newcount\hour
\newcount\minute
\newtoks\amorpm
\hour=\time\divide\hour by60
\minute=\time{\multiply\hour by60 \global\advance\minute by-\hour}
\edef\standardtime{{\ifnum\hour<12 \global\amorpm={am}%
	\else\global\amorpm={pm}\advance\hour by-12 \fi
	\ifnum\hour=0 \hour=12 \fi
	\number\hour:\ifnum\minute<10 0\fi\number\minute\the\amorpm}}
\edef\militarytime{\number\hour:\ifnum\minute<10 0\fi\number\minute}
\def\draftlabel#1{{\@bsphack\if@filesw {\let\thepage\relax
   \xdef\@gtempa{\write\@auxout{\string
      \newlabel{#1}{{\@currentlabel}{\thepage}}}}}\@gtempa
   \if@nobreak \ifvmode\nobreak\fi\fi\fi\@esphack}
	\gdef\@eqnlabel{#1}}
\def\@eqnlabel{}
\def\@vacuum{}
\def\draftmarginnote#1{\marginpar{\raggedright\scriptsize\tt#1}}
\def\draft{\oddsidemargin -.5truein
	\def\@oddfoot{\sl preliminary draft \hfil
	\rm\thepage\hfil\sl\today\quad\militarytime}
	\let\@evenfoot\@oddfoot	\overfullrule 3pt
	\let\label=\draftlabel
	\let\marginnote=\draftmarginnote
   \def\@eqnnum{(\theequation)\rlap{\kern\marginparsep\tt\@eqnlabel}%
\global\let\@eqnlabel\@vacuum}  }
\def\preprint{\twocolumn\sloppy\flushbottom\parindent 2em
	\leftmargini 2em\leftmarginv .5em\leftmarginvi .5em
	\oddsidemargin -.5in	\evensidemargin -.5in
	\columnsep .4in	\footheight 0pt
	\textwidth 10in	\topmargin  -.4in
	\headheight 12pt \topskip .4in
	\textheight 7.1in \footskip 0pt
	\def\@oddhead{\thepage\hfil\addtocounter{page}{1}\thepage}
	\let\@evenhead\@oddhead	\def\@oddfoot{}	\def\@evenfoot{} }
\def\numberbysection{\@addtoreset{equation}{section}
	\def\theequation{\thesection.\arabic{equation}}}
\def\underline#1{\relax\ifmmode\@@underline#1\else
	$\@@underline{\hbox{#1}}$\relax\fi}
\def\titlepage{\@restonecolfalse\if@twocolumn\@restonecoltrue\onecolumn
     \else \newpage \fi \thispagestyle{empty}\c@page\z@
	\def\thefootnote{\fnsymbol{footnote}} }
\def\endtitlepage{\if@restonecol\twocolumn \else \newpage \fi
	\def\thefootnote{\arabic{footnote}}
	\setcounter{footnote}{0}}  
\catcode`@=12
\relax
\def\figcap{\section*{Figure Captions\markboth
	{FIGURECAPTIONS}{FIGURECAPTIONS}}\list
	{Figure \arabic{enumi}:\hfill}{\settowidth\labelwidth{Figure 999:}
	\leftmargin\labelwidth
	\advance\leftmargin\labelsep\usecounter{enumi}}}
 \relax
\def\tablecap{\section*{Table Captions\markboth
	{TABLECAPTIONS}{TABLECAPTIONS}}\list
	{Table \arabic{enumi}:\hfill}{\settowidth\labelwidth{Table 999:}
	\leftmargin\labelwidth
	\advance\leftmargin\labelsep\usecounter{enumi}}}
 \relax
\def\reflist{\section*{References\markboth
	{REFLIST}{REFLIST}}\list
	{[\arabic{enumi}]\hfill}{\settowidth\labelwidth{[999]}
	\leftmargin\labelwidth
	\advance\leftmargin\labelsep\usecounter{enumi}}}
 \relax
\makeatletter
\newcounter{pubctr}
\def\publist{\@ifnextchar[{\@publist}{\@@publist}}
\def\@publist[#1]{\list
	{[\arabic{pubctr}]\hfill}{\settowidth\labelwidth{[999]}
	\leftmargin\labelwidth
	\advance\leftmargin\labelsep
	\@nmbrlisttrue\def\@listctr{pubctr}
	\setcounter{pubctr}{#1}\addtocounter{pubctr}{-1}}}
\def\@@publist{\list
	{[\arabic{pubctr}]\hfill}{\settowidth\labelwidth{[999]}
	\leftmargin\labelwidth
	\advance\leftmargin\labelsep
	\@nmbrlisttrue\def\@listctr{pubctr}}}
 \relax
\makeatother
\catcode`\@=11
\def\section{\@startsection {section}{1}{0pt}{-3.5ex plus -1ex minus
 -.2ex}{2.3ex plus .2ex}{\raggedright\large\bf}}
\catcode`\@=12
\newskip\humongous \humongous=0pt plus 1000pt minus 1000pt

\newif\ifdtup

\def\oldreffmt#1{\rlap{[#1]} \hbox to 2\parindent{}}

\def\figfmt#1{\rlap{Figure {#1}} \hbox to 1in{}}

\let\vev\VEV

\def\abs#1{\left| #1\right|}

\def\half{{1\over 2}}
\def\beq{\begin{equation}}
\def\eeq{\end{equation}}

\def\bea{\begin{eqnarray}}

\def\eea{\end{eqnarray}}
\def\np#1#2#3{           {\it Nucl. Phys. }{\bf #1}, #2 (19#3)}
\def\pl#1#2#3{           {\it Phys. Lett. }{\bf #1}, #2 (19#3)}
\def\pr#1#2#3{           {\it Phys. Rev. }{\bf #1}, #2 (19#3)}

\def\prl#1#2#3{          {\it Phys. Rev. Lett. }{\bf #1}, #2 (19#3)}

\def\zp#1#2#3{           {\it Zeit. fur Physik }{\bf #1}, #2 (19#3)}
\catcode`\@=11
\def\eqnarray{\stepcounter{equation}\let\@currentlabel=\theequation
\global\@eqnswtrue
\global\@eqcnt\z@\tabskip\@centering\let\\=\@eqncr
\gdef\@@fix{}\def\eqno##1{\gdef\@@fix{##1}}%
$$\halign to \displaywidth\bgroup\@eqnsel\hskip\@centering
  $\displaystyle\tabskip\z@{##}$&\global\@eqcnt\@ne
  \hskip 2\arraycolsep \hfil${##}$\hfil
  &\global\@eqcnt\tw@ \hskip 2\arraycolsep $\displaystyle\tabskip\z@{##}$\hfil
   \tabskip\@centering&\llap{##}\tabskip\z@\cr}

\def\@@eqncr{\let\@tempa\relax
    \ifcase\@eqcnt \def\@tempa{& & &}\or \def\@tempa{& &}
      \else \def\@tempa{&}\fi
     \@tempa \if@eqnsw\@eqnnum\stepcounter{equation}\else\@@fix\gdef\@@fix{}\fi
     \global\@eqnswtrue\global\@eqcnt\z@\cr}

\catcode`\@=12
\font\tenbifull=cmmib10 
\font\tenbimed=cmmib10 scaled 800
\font\tenbismall=cmmib10 scaled 666
\relax

\def\thefootnote{\fnsymbol{footnote}}
\newcommand{\ba}{\begin{eqnarray*}}
\newcommand{\ea}{\end{eqnarray*}}
\newcommand{\Nbar}{\overline{N}}
\newcommand{\ux}{{$U(1)_X\,$}}
\newcommand{\sm}{{$SU(3)_c\times SU(2)_L \times U(1)_Y\,$}}
\newcommand{\ew}{{$SU(2)_L \times U(1)_Y\,$}}
\newcommand{\md}{m_{\Delta}^2}
\newcommand{\ms}{m_{\Sigma}^2}
\newcommand{\nr}{\nu_R}
\newcommand{\snr}{{\tilde{\nu}}_{R}}
\newcommand{\mn}{m_{N}^2}
\newcommand{\mnbar}{m_{\Nbar}^2}
\newcommand{\mhd}{m_{H_1}^2}
\newcommand{\mhu}{m_{H_2}^2}
\newcommand{\gsim}{\,\raisebox{-.5ex}{\rlap{$\sim$}}\raisebox{.5ex}{$>$}\,}
\newcommand{\fb}{\overline{5}}
\begin{document}
\begin{titlepage}
\begin{center}
\today     \hfill    LBL-35950 \\
          \hfill    UCB-PTH-94/21 \\
          \hfill    hep-ph/9411276 \\

\vskip .3in

{\large \bf Squark and Slepton Mass Relations In Grand Unified Theories}
\footnote{This work was supported in part by the Director, Office of
Energy Research, Office of High Energy and Nuclear Physics, Division of
High Energy Physics of the U.S. Department of Energy under Contract
DE-AC03-76SF00098 and in part by the National Science Foundation under
grant PHY-90-21139.}

\vskip .3in
H.-C. Cheng
and
L.J. Hall

\vskip .1in

{\em Theoretical Physics Group\\
    Lawrence Berkeley Laboratory\\
      and\\
    Department of Physics\\
      University of California\\
    Berkeley, California 94720}
\end{center}

\vskip .3in

\begin{abstract}
In the minimal supersymmetric standard model, assuming universal scalar
masses  at large energies, there are four intragenerational relations
between the masses of the squarks and sleptons for each light generation.
In this paper we study the scalar mass relations which follow only
from the assumption that at large energies there is a grand unified theory
which leads to a significant prediction of the weak mixing angle.
Two new intragenerational mass relations for each of the light generations
are derived. In addition, a third mass relation is found which relates the
Higgs masses, the masses of the third generation scalars , and the masses
of the scalars of the lighter generations. Verification of a fourth mass
relation, involving
only the charged slepton masses, provides a signal for SO(10) unification.
\end{abstract}

\end{titlepage}
\renewcommand{\thepage}{\roman{page}}
\setcounter{page}{2}
\mbox{ }

\vskip 1in

\begin{center}
{\bf Disclaimer}
\end{center}

\vskip .2in

\begin{scriptsize}
\begin{quotation}
This document was prepared as an account of work sponsored by the United
States Government. While this document is believed to contain correct
 information, neither the United States Government nor any agency
thereof, nor The Regents of the University of California, nor any of their
employees, makes any warranty, express or implied, or assumes any legal
liability or responsibility for the accuracy, completeness, or usefulness
of any information, apparatus, product, or process disclosed, or represents
that its use would not infringe privately owned rights.  Reference herein
to any specific commercial products process, or service by its trade name,
trademark, manufacturer, or otherwise, does not necessarily constitute or
imply its endorsement, recommendation, or favoring by the United States
Government or any agency thereof, or The Regents of the University of
California.  The views and opinions of authors expressed herein do not
necessarily state or reflect those of the United States Government or any
agency thereof or The Regents of the University of California and shall
not be used for advertising or product endorsement purposes.
\end{quotation}
\end{scriptsize}

\vskip 2in

\begin{center}
\begin{small}
{\it Lawrence Berkeley Laboratory is an equal opportunity employer.}
\end{small}
\end{center}

\newpage
\renewcommand{\thepage}{\arabic{page}}
\setcounter{page}{1}

\section*{I. Introduction}

If supersymmetry is a symmetry of nature, broken only at the weak scale, then
future experiments will discover many extra particles, in particular the
superpartners of all the quarks and leptons.
The masses of these scalar quarks and leptons will provide extra clues about a
more fundamental theory at higher energies.
However, whereas the quark and lepton masses provide information on how chiral
and flavor symmetries are broken, the squark and slepton masses will provide a
window to the structure of supersymmetry breaking.

It may be that the squark and slepton  spectrum will show no clear pattern or
regularities, and the origin of the spectrum will become a major puzzle, rather
like the present situation with quark and lepton masses.
However, much attention
has been focussed on a single theory, the minimal supersymmetric standard model
(MSSM), in which a very clear pattern emerges in the scalar spectrum. By the
MSSM we will mean the supersymmetric extension of the standard model with
minimal field content, which has a boundary condition near the Planck scale
that
the soft supersymmetry breaking mass parameters for the scalars are all equal.
In
this model, the physical masses of the 14 squarks and sleptons of the lighter
two generations are given in terms of just 5 unknown parameters: the universal
scalar masses at the Planck scale, $m^2_0$, the three gaugino masses, $M_a$,
and
the ratio of electroweak breaking vevs, $\tan\beta = v_2/v_1$.
Due to effects of large Yukawa couplings, the physical squark and slepton
masses
 of the heaviest generation depend on one further parameter, $A$.
Although these effects are well understood and can easily be added, for
simplicity, we consider only the lightest two generations.
Thus the MSSM has many relations amongst the scalar masses.
However, the question as to why all scalars are assumed degenerate at the
Planck
scale becomes extremely important. If experiments are done to check the
validity
of the scalar mass relations of the MSSM \cite{exp},
what is the fundamental principle
which
is being tested?

Flavor changing processes provide considerable experimental constraints
on the
form of the squark and slepton mass matrices \cite{dg,hkt}.
However, these constraints are intimately connected with flavor violation
and provide constraints between the masses of scalars of different generations.
For a given generation there are 5 independent gauge invariant squark and
slepton masses: $m_Q, m_{U^c}, m_{D^c}, m_L$ and $m_{E^c}$, where $Q$ and $L$
represent $SU(2)$ doublet squarks and sleptons, while $U,D$ and $E$ are
$SU(2)$
singlet squarks and sleptons. Certainly the flavor changing constraints do not
constrain the ratios $m_Q: m_U : m_D : m_L : m_E$, and it is largely these
ratios which will be addressed in this paper.

The assumption of a universal scalar mass at high energies originated from
studies of $N=1$ supergravity theories in which supersymmetry is broken in a
hidden sector. The scalar mass was found to be universal in particular models
\cite{can,bfs} and also in a wide class of models \cite{hlw}.
However, the universal mass is not
a general property of supergravity models, and involves an assumption about the
form of the K\"ahler potential. If there are $N$ fields in the observable
sector
of the theory, an $SU(N)$ invariance of the K\"ahler potential guarantees the
universality of the scalar masses at the Planck scale \cite{hlw}.
However, this symmetry is clearly broken elsewhere in the theory, and so the
universality of the scalar masses can only be understood as a special property
of certain supergravity theories. If the scalar mass relations of the MSSM were
violated, it might simply mean that the K\"ahler potential does not possess
this $SU(N)$ invariance.

In this paper we study squark and slepton mass relations which follow from two
assumptions, which have nothing to do with supergravity.

(1) The standard model is unified into a grand unified theory (GUT).

It is well know that a grand unified symmetry, together with supersymmetry,
has
yielded a successful relation amongst the gauge couplings of the standard
model \cite{guni}.
Much attention has also been given to quark and lepton mass relations which can
follow from a grand unified symmetry. It therefore seems well worthwhile
studying what squark and slepton mass relations might follow purely from grand
unification.

(2) The generation changing entries in the squark and slepton
masses (in a basis
   where the quark and lepton masses are diagonal) are sufficiently small not
to
   affect the scalar mass  eigenvalues at a level of accuracy to which the mass
   relations will be experimentally tested.

In fact, the latter is hardly an assumption, such large flavor changing effects
are almost certainly experimentally excluded.
Since the grand unified symmetry acts within a generation, we expect relations
amongst squark and slepton masses of the same generation, we do not expect any
relations between masses of particles in different generations.

We begin section II by writing down the mass relations between squarks and
sleptons of a given generation which occur in the MSSM. We then list the
assumptions which a supersymmetric grand unified theory
(SGUT) must satisfy for a successful weak mixing angle
prediction to occur at the $1\%$ level. Finally, we show that, with these
assumptions, we are able to derive
two intrageneration scalar mass relations.
The mass relation of the MSSM which
relates the masses of the two charged sleptons within a generation
may be violated. This is
a particularly important mass relation since it is likely that the squarks
will be much heavier than the sleptons, and this will be the first mass
relation of the MSSM to be tested. In section III we study the extent to
which this mass relation is expected to follow if the GUT gauge groups
includes $SO(10)$. While this slepton mass relation is generically expected
as a consequence of the $SO(10)$ gauge symmetry, we find that
radiative corrections and additional
$D$-term contributions to the scalar masses, beyond those of the MSSM,
may lead to its violation. In section IV we show that even if the additional
$D$-term contributions do not arise at tree level, they could be generated
by radiative corrections.
In section V we show that these extra $D^2$
interactions found in $SO(10)$ could lead to an easing of the fine tunning
problem which has been found when the MSSM has large $\tan \beta$ and the
universal scalar mass boundary condition.

\section*{II. Scalar Mass Relations In A Class of Grand Unified Theories}

Before studying grand unified theories, we give the well known predictions for
the scalar masses in the MSSM, taken to have universal scalar masses $m^2_0$
at the Planck scale.
Mass splittings arise from renormalization group scaling from
Planck to weak scales \cite{rg}, and
the renormalization group equations are given by
\ba
{d \over d \ln \mu} m_i^2(\mu) & = & {1\over 16\pi^2}[-8C_2(R_a^i) g_a^2(\mu)
M_a^2(\mu) +{6\over 5} Y_i g_1^2(\mu) S(\mu) \samepage\\ \samepage
& & + \sum_{j,k} \abs{\lambda_{ijk}}^2
(m_i^2 +m_j^2 +m_k^2 + A_{ijk}^2)] , \eqno{(2.1)} \samepage\\
\samepage
{d \over d \ln \mu} S(\mu) & = & {b_1 \over 2\pi} \alpha_1(\mu) S(\mu) ,
\eqno{(2.2)} \samepage\\  \samepage
S(\mu) & = & \sum_i Y_i m_i^2(\mu) , \eqno{(2.3)}
\ea
where $a=1,2,3$ represents $U(1)_Y,\, SU(2)_L $ and $SU(3)_c$,\footnote{The
$SU(5)$ GUT normalization, $g_1^2 = {5\over 3}{g'}^2$, is used for the
$U(1)$ coupling.}
$i$ represents the species of the scalar and $Y_i$ is the corresponding
hypercharge,
$A_{ijk}$'s are the soft SUSY breaking trilinear scalar couplings,
and $\lambda_{ijk}$'s are the superpotential couplings.
$C_2(R_a^i)$ is the second Casimir invariant of the gauge group $a$
for the species $i$, $C_2 = {N^2-1 \over 2N}$ for the fundamental
representation of $SU(N)$, ${3\over 5}Y_i^2$ for $U(1)_Y$.
The $S$ term is zero under the assumption of universal scalar masses
and hence does not contribute.
For the lightest two generations, whose superpotential coupling contributions
are negligible, the mass splittings involve
only contributions from the gauginos, which have masses $M_{0a}$ at the Planck
scale.
Mass splittings also arise from the $D^2$ terms of the potential due to
\ew  interactions.
These are proportional to $M^2_Z \cos 2\beta$.
The result is
$$
m^2_i(\mu) = m^2_0 +\sum_a f_{ai} M^2_{0a} + (T_{3_i} - Q_i \sin^2 \theta_W)
M_Z^2 \cos 2\beta, \eqno(2.4)
$$
where $i$ runs over the 7 types of squark and slepton: $U,D,U^c,D^c,E,N$ and
$E^c$, and it is understood that the two light generations have identical
scalar
spectra. The renormalization constants $f_{ai}$ are
$$
f_{ai} (\mu) = {2\over b_a} C_2(R^i_a)\left({\alpha^2_a (\mu)\over
\alpha^2_a
(M_p)} -1\right). \eqno(2.5)
$$
where $b_a$ is the 1-loop beta function coefficient,
and $\mu$ should be
taken equal to the scalar mass, $m_i$.

Suppose that $\beta$ is known, for example from a Higgs mass measurement,
then
the 7 values of $m^2_i$ depend only on 4 unknown parameters, $m_0$ and $M_{0a}$
yielding three intragenerational mass relations for the MSSM \cite{mp}.
Two further relations follow if $M_{0a}$ is independent of $a$.
In the following the scalar masses are scaled to the same renormalization
point so that these mass relations can be displayed in simpler forms,

Two of these relations have only to do with $SU(2)$ breaking and are
$$
m^2_U - m^2_D = m^2_N - m^2_E = M^2_Z \cos 2\beta \cos^2 \theta_W. \eqno(2.6)
$$
These splittings arise because of the differing $T_3$ quantum numbers of
the
upper and lower components of the doublets $Q = (U,D)$ and $L = (N,E)$.
It is convenient to define $m^2_Q$ and $m^2_L$ as the average squared mass
of
the doublet representation, thus
$m^2_Q = \half (m^2_U + m^2_D)$ and $m^2_L = \half (m^2_N + m^2_E)$.
In the rest of this paper it is the masses $m^2_I, I = 1 ... 5$ of the five
types of
multiplet $Q, U^c, D^c L,$ and $E^c$ which will interest us.
In the MSSM, these are:
$$
m^2_I = m^2_0 + \sum_a f_{aI} M^2_{0a} - Y_I \sin^2 \theta_W M_Z^2 \cos
2\beta, \eqno(2.7)
$$
where $Y_I$ is the hypercharge of multiplet $I$ ($Q = T_3 + Y)$.

The mass predictions of (2.7) are based on several strong assumptions.
The universal scalar mass is a speculative assumption about the form of the
interactions in supergravity, and has been questioned, particularly by those
working on string-inspired models \cite{string}.
The mass formula of equation (2.4) assumes the minimal particle content beneath
the
Planck scale. If there are extra gauge interactions then the index a $=
1,2,3,4...,$ yielding extra terms.
If there are extra chiral fields with gauge quantum number then the $b_a$ of
equation (2.5) will change.
Furthermore, if these extra chiral fields allow further superpotential
interactions of strength $\lambda$ involving quark and lepton fields, then
additional
terms proportional to $\lambda^2$ will contribute to $m^2_i(\mu)$.

In this paper we study the scalar mass relations which follow from certain
assumptions about grand unification.
The assumptions appear to us to be better motivated than those listed above
for
the MSSM, since they are based on the successful supersymmetric GUT
prediction
for $\sin^2\theta_W$ \cite{guni},
 the weak mixing angle, which we briefly summarize.
The combined fit to the precision LEP data gives $\sin^2\theta_W (LEP) = 0.2321
\pm
0.0005$, which corresponds to $m_t = 176 \pm$ 15 GeV (these results,
and the results of
other fits to experimental data given below, are taken from \cite{erler}).
The left right asymmetry measurement at SLAC gives: $\sin^2\theta_W$ (SLAC)
 = 0.2292 $\pm$ 0.0010 and $m_t = 255^{+22}_{-24}$ GeV.
The $W$ mass measurements from CDF, D0 and UA2 correspond to $\sin^2
\theta_W = 0.2326 \pm 0.0008$.
These experimental numbers should be compared with the supersymmetric GUT
central prediction of $\sin^2\theta_W (SGUT) = 0.2342 \pm 0.0014$, where the
only uncertainty shown is that due to $\alpha_s(M_Z) = 0.120 \mp 0.005$.
In addition, simple models could have uncertainties of 0.0030 from
threshold
corrections at the GUT and weak scales.
The weak mixing angle therefore provides the only successful theoretical
prediction at the 1\% level of any parameter of the standard model.
This suggests that we take the assumptions which are {\em sufficient}
to get this prediction and
use them to make predictions for the squark and slepton masses.
These assumptions are

1. At some scale $M_G$ the gauge group is $SU(5) \times G$, where $SU(5)$
   contains the entire standard model gauge group.

2. At mass scales below $M_G$ the gauge group is $SU(3)_c \times SU(2)_L \times
U(1)_Y \times G'$.

3. At mass scales below $M_G$ the only particles coupling to the standard
model  gauge interactions are those of the MSSM.\footnote{In fact the
prediction
 of $\sin^2\theta_W$ is not altered if extra complete, degenerate $SU(5)$
   multiplets occur beneath $M_G$.
We assume these to be absent; it could be worth studying the extent to which
such representations affect the scalar mass relations. }

These assumptions are not a necessary requirement for an acceptable value
of $\sin^2 \theta_W$. Acceptable values can be obtained in very many ways,
for example in non-supersymmetric $SU(5)$ theories with extra multiplets
which are not $SU(5)$ degenerate \cite{my}.
However, it is these assumptions which
uniquely produce a significant prediction. All the other schemes have a
free parameter which can be chosen to fit
$\sin^2 \theta_W$.\footnote{
In the MSSM the scale of supersymmetry breaking is not a free parameter -
it is determined to be of order the weak scale by radiative
electroweak symmetry breaking.}

What scalar mass relations follow from these assumptions?
The first assumption imposes the boundary condition (which is taken to be at
$M_G$ now) on scalar masses within the same generation:

$$
m_{Q_0} = m_{E^c_0} = m_{U^c_{0}} = m_{10},\eqno(2.8)
$$
$$
m_{L_0} = m_{D^c_0} = m_{\overline{5}},\eqno(2.9)
$$
because $Q,E^c$ and $U^c$ all lie in a 10 dimensional representation, and
$L$ and $D^c$ lie in the $\overline{5}$.
There is no boundary condition relating masses of particles in different
generations, and hence no such mass relations will result.

Let us study a particular generation, and suppose that in the $SU(5)\times G$
theory it lies in representation $(10, R_1) + (\overline{5}, R_2)$.
If $R_1$ and $R_2$ are non-trivial and if $G$ breaks to $G'$ which is
non-trivial, then the $G'$ gauginos can renormalize the squark and slepton
masses. However, since all members of the 10 have the same $G'$ quantum
numbers,
this renormalization is common, and can simply be absorbed into the unknown
parameter $m_{10}$.
An identical situation applies to the $\overline{5}$.
Hence the common mass $m_0^2$ in
the formula (2.7) should be replaced by
$m^2_0 \to m^2_{I_0}$ which take on the two possible
values shown in (2.8) and (2.9) according to whether $I$ lies in a 10 or
$\overline{5}$ representation.
In addition, the $S$ term, which vanishes under the universal boundary
condition assumption, is now given by
$$
S(M_G) = \sum_i Y_i m_i^2(M_G) = m_{H_2}^2(M_G) - m_{H_1}^2(M_G). \eqno(2.10)
$$
Since $H_2$ and $H_1$ lie in different representations of $SU(5)$,
$m_{H_2}^2(M_G)$ and $m_{H_1}^2(M_G)$ are not necessarily equal.
{}From (2.2) it follows that $S$ scales as $\alpha_1$,
$$
{S(\mu) \over S(\mu_0)} = {\alpha_1(\mu) \over \alpha_1(\mu_0)}. \eqno(2.11)
$$
The contributions of the $S$ term can be written as
$$
\delta_S m_i^2(\mu) = Y_i\, T, \eqno(2.12)
$$
where
\ba
T &= - {3 \over 5b_1} (S(M_G)-S(\mu)) = -{3 \over 5b_1}S(M_G)
(1- {\alpha_1(\mu) \over \alpha_1(M_G)}) \samepage \\
 &= -{3 \over 5b_1}S(\mu)
({\alpha_1(M_G) \over \alpha_1(\mu)}-1). \eqno{(2.13)}
\ea
Among the 5 masses (2.7) of each light generation, there are 3 combinations
independent of $m_{I_0}^2$:
\ba
m^2_Q - m^2_{U^c} &=& (C_2 -{15\over36}C_1) M^2_0 - {5\over 6} \sin^2\theta_W
M^2_Z \cos 2 \beta + {5\over 6}T  , \eqno{(2.14a)} \samepage \\
m^2_Q - m^2_{E^c} &=& (C_3 + C_2 - {35\over 36} C_1) M^2_0 + {5\over 6}
\sin^2\theta_W M^2_Z \cos 2\beta -{5\over 6}T , \eqno{(2.14b)} \samepage \\
m^2_{D^c} - m^2_L &=& (C_3 - C_2 - {5\over 36} C_1) M^2_0 - {5\over 6} \sin^2
\theta_W M^2_Z \cos 2\beta +{5\over 6}T , \eqno{(2.14c)}
\ea
where we have written $f_{3i} = C_3$ for a color triplet, $f_{2i} = C_2$ for a
weak doublet and $f_{1i}= Y^2_i C_1$, and the $\alpha_a(M_p)$ in $f_{ai}$
should be replaced by $\alpha_a(M_G)$.
By rearranging the above equations,
we arrive at the following two mass relations independent of $T$:
\ba
2m_Q^2 - m_{U^c}^2 - m_{E^c}^2 &=& (C_3 + 2C_2 - {25 \over 18}C_1)M_0^2,
\eqno{(2.15a)} \samepage \\
m_Q^2 + m_{D^c}^2 - m_{E^c}^2 - m_L^2 &=& (2C_3 - {10\over 9}C_1)M_0^2,
\eqno{(2.15b)}
\ea
and also an expression for $T$:
\ba
T = {3\over 10}
(m_Q^2 - 2m_{U^c}^2 + m_{D^c}^2 +m_{E^c}^2 - m_L^2 + {10\over 3}
\sin^2 \theta_W M_Z^2 \cos 2\beta) .\eqno{(2.16)}
\ea
Since $S(M_G)$ is only proportional to the difference
$m_{H_2}^2(M_G) - m_{H_1}^2(M_G)$ and $b_1={33\over 5}$,
we have $|T|< {1\over 11}|m_{H_2}^2(M_G) - m_{H_1}^2(M_G)|$.
If the splitting between $m_{H_2}^2(M_G)$ and $ m_{H_1}^2(M_G)$ is not
too large, then
$T$ is small and the these mass relations of (2.14), with $T=0$, are
approximately true.
Alternatively, one can use (2.3), (2.13) and (2.16) to get
\ba
&(m_Q^2 - 2m_{U^c}^2 + m_{D^c}^2 +m_{E^c}^2 - m_L^2
 + {10\over 3}\sin^2 \theta_W M_Z^2 \cos 2\beta
)_{\mbox{\scriptsize \it 3rd generation}}
\samepage \\
& \hspace{-0.4in} + (m_{H_2}^2-m_{H_1}^2)\, = \, S(\mu)-{20\over 3}T
\,= \,- ({11\alpha_1(\mu) \over \alpha_1(M_G) - \alpha_1(\mu)})\,T
-{20\over 3}\,T
\samepage \\
& = -({20\alpha_1(M_G) +13\alpha_1(\mu) \over 3\alpha_1(M_G)- 3\alpha_1(\mu)})
\,T. \eqno{(2.17)}
\ea
This combination does not suffer  from the renormalization effects of
the large third generation Yukawa couplings,
Using $T$ from (2.16) in (2.17) gives a third (intergeneration)
mass relation:
\ba
& (m_Q^2 - 2m_{U^c}^2 + m_{D^c}^2 +m_{E^c}^2 - m_L^2
 + {10\over 3}\sin^2 \theta_W M_Z^2 \cos 2\beta
)_{\mbox{\scriptsize \it 1st or 2nd gen.}}
\samepage \\
=& -\{ (m_Q^2 - 2m_{U^c}^2 + m_{D^c}^2 +m_{E^c}^2 - m_L^2
 + {10\over 3}\sin^2 \theta_W M_Z^2 \cos 2\beta
)_{\mbox{\scriptsize \it 3rd gen.}}
\samepage \\
& \hspace{-1.5in} + (m_{H_2}^2-m_{H_1}^2) \} \times
{10\alpha_1(M_G)-10\alpha_1(\mu) \over 20\alpha_1(M_G) +13\alpha_1(\mu)}.
\eqno{(2.18)}
\ea

The MSSM provides 4 mass relations within each generation: those of (2.14)
with $T=0$ together with
$$
m^2_L- m^2_{E^c} = (C_2 - {3\over 4} C_1) M^2_0 + {3\over 2} \sin^2
\theta_W M^2_Z \cos 2\beta, \eqno(2.19)
$$
and also predicts identical spectra for each of the light generations.

In this section we have shown that two of these mass relations follow from
a completely different boundary condition assumption than the one of
universal scalar masses used for the MSSM. We have found that, in any GUT
where the successful prediction of the weak mixing angle at the $1\%$
accuracy level is preserved, 2 of the 4 mass relations of the MSSM for
each light generation is preserved and a third one can be recovered
provided that
the third generation scalar masses and Higgs masses are also measured.


\section*{III. An Extra Mass Relation in SO(10)?}

The mass relation (2.19) can be reformulated as a relation between the two
charged slepton masses of a given generation:
$$
m_E^2 - m_{E^c}^2 = (C_2 - {3\over 4}C_1)M_0^2 + (-{1\over 2} + 2 \sin^2
\theta_W)M_Z^2 \cos 2\beta. \eqno{(3.1)}
$$
In the following we will not include the contributions from the $S$ term,
it is assumed to be small or can be obtained from (2.16) or (2.17), then be
substracted from the scalar masses.
This relation is particularly important because:

(a) The super-QCD interactions tend to increase the  masses of the squarks
above the sleptons, hence we expect this to be the first scalar mass
relation of the MSSM to be tested.

(b) We have shown that this relation is precisely the one which cannot be
deduced from $SU(5)$ unification. This is clearly because $E$ and $E^c$ are
in different representations of $SU(5)$.

If the gauge group is extended to include $SO(10)$, such that a single
generation lies entirely in a 16 dimensional spinor representation,
then it is tempting to think that this slepton mass relation will be
recovered, perhaps one can view this particular mass relation as a low
energy signature of $SO(10)$. In this section, we explore in more detail
the extent  to which this is true.

We will make the three assumptions, given in the last section, necessary
for the GUT to yield a significant $\sin^2 \theta_W$ prediction. In addition
we add a 4th assumption:

4. At energy scales greater than $M_{10}$, which is greater than or equal to
$M_G$, the gauge group contains a factor  which includes the usual $SO(10)$
gauge group.

This assumption provides the extra boundary condition which sets
$m_L(\mu)$ and $m_{E^c}(\mu)$ equal at $\mu \ge M_{10}$. The crucial
question now is: are there any additional effects which could split
these masses other than those of the \ew  gaugino contributions and the
\ew $D^2$ interactions, shown in (2.19) and (3.1)?

There are 4 such effects , which could break the slepton mass relation
in an important way \cite{hkr,kmy,pp}:

(a) Radiative contributions from the gauge couplings and gaugino masses
between $M_{10}$ and $M_G$,

(b) Radiative contributions from the superpotential couplings between
$M_{10}$ and $M_G$,

(c) Tree level $D$-term  contributions,

(d) Radiatively generated $D$-term contributions.

Suppose that $M_{10}$ is higher than $M_G$, and that
beneath $M_{10}$ $SO(10)$ breaks down to $SU(5)$ (or $SU(5) \times U(1)_X$).
The two charged sleptons of a given generation belong to $\overline{5}$ and
10 representations of $SU(5)$ respectively and therefore their masses receive
different radiative corrections. The radiative correction contributions
from the $SU(5)$ gaugino mass is
$$
\delta m^2(R) = {2\over b_5} C_2(R) (1- {\alpha_5^2(M_{10}) \over
\alpha_5^2(M_G)}) M_5^2(M_G), \eqno{(3.2)}
$$
where $C_2(\overline{5})={12 \over 5}$ and $C_2(10)={18 \over 5}$.
Therefore we have ${\delta m^2(10)\over \delta m^2(\overline{5})} ={3\over 2}$.
If \ux survives beneath $M_{10}$, the \ux gaugino mass also contributes to
the radiative corrections and reduces this ratio
($X_{10}=-1,\; X_{\overline{5}}
=3$), but in general its contributions are smaller.

If this is the only source which violate the slepton mass relation, then we
have
$$
1 \le {m_{10}\over m_{\fb}} \le {3 \over 2}, \eqno{(3.3)}
$$
and the violation should be small if gaugino mass is found to be small unless
the gauge coupling increases very rapidly above $M_G$.

In addition to the radiative corrections from the gauge couplings, if the
sleptons have some superpotential coupling of strength $\lambda$ with
fields which acquire masses ${\cal O}(M_G)$, then there are radiative
corrections to the slepton masses between $M_{10}$ and $M_G$ at order
$\lambda^2$. In order to generate significant  violations of the slepton
mass relation, $\lambda$ has to be large, probably $\gsim {1\over 3}$,
but such a large superpotential coupling could also destroy the degeneracy
of scalar masses of different generations and induce unacceptable flavor
changing
effects unless there is a horizontal symmetry above $M_G$ which keeps
the scalar masses  of the two lighter generations degenerate.

$D$-term contributions to scalar masses can arise when the rank of the gauge
group is reduced. To see this, consider the following situation.
Suppose the \ux subgroup of $SO(10)$ ($SO(10) \supset SU(5) \times U(1)_X$)
is broken by the vacuum expectation values (VEVs) of $N$ and $\Nbar$ fields
which lie in 16 and $\overline{16}$ representations of $SO(10)$. The \ux
gauge interaction contains a piece
$$
{1 \over 2} g_X^2 (X_N \abs{N}^2 - X_N |{\Nbar}|^2 +
    \sum_i X_i \abs{\phi_i}^2)^2, \eqno{(3.4)}
$$
where $X_i$ is the X charge of the $\phi_i$ field. When the VEVs of $N$ and
$\Nbar$ fields are not equal, it gives extra contributions to the squared
masses of scalar fields of nonzero X charges. This happens if
the soft SUSY breaking masses of $N$ and $\Nbar$ are different
\cite{kmy,hk,drees}.
The relevant part of the
scalar potential for these fields  we take to be
\ba
V(N,\, \Nbar) & = & {1 \over 2} g_X^2 (X_N \abs{N}^2 - X_N |{\Nbar}^2|)^2
           \samepage\\
    & &   + \mn \abs{N}^2 + \mnbar |{\Nbar}^2| + \lambda^2|N\Nbar-\mu^2|^2,
     \eqno{(3.5)}
\ea
where $\mn$ and $\mnbar$ are the soft SUSY breaking masses of the $N$ and
$\Nbar$ fields, and they are of the order of the SUSY breaking scale $m_S$.
The last term is to give large VEVs ($\simeq \mu$) to $N$ and $\Nbar$
fields.\footnote{Different ways of stablizing the VEVs of $N$ and $\Nbar$ do
not change the basic result, they only give corrections to the higher order
terms in equation (3.7).}
Defining $\Sigma \equiv \abs{N}^2 + |{\Nbar}^2|$, $\Delta \equiv
\abs{N}^2 - |{\Nbar}^2|$, $\ms \equiv {1 \over 2}(\mn + \mnbar)$ and
$\md \equiv {1 \over 2}(\mn - \mnbar)$, we can rewrite $V$ as
$$
V  =  {1 \over 2} g_X^2 (X_N \Delta )^2
    + \ms \Sigma + \md \Delta + \lambda^2 |{1\over 2}\sqrt{\Sigma^2-\Delta^2}
    -\mu^2|^2. \eqno{(3.6)}
$$
Minimizing the potential with respect to $\Delta$
we obtain
$$
\Delta = -{m_{\Delta}^2 \over X_N^2 g_X^2} + {\cal O}({m_S^3\over \mu})
, \eqno{(3.7)}
$$
This shifts the mass of the scalar particle with charge
$X_i$ by the amount
$$
\delta m_i^2 = g_X^2 X_i X_N \Delta \simeq - {X_i \over X_N} \md . \eqno{(3.8)}
$$
Therefore any scalar particle which carries \ux charge will receive a tree
level $D$-term contribution which is proportional to its \ux charge and
the difference
of the soft-breaking masses $\mn$ and $\mnbar$. Since $N$ and $\Nbar$ lie in
different representations of $SO(10)$, $SO(10)$ allows $\mn$ to be
very different from $\mnbar$, and also $X_{10}$ and $X_{\fb}$ are different
($X_{10}=-1,\; X_{\fb}=3$),
this provides a large breaking of the slepton relation (2.19), (3.1).

{}From the above discussion it follows that a significant violation of the
slepton mass relation by the $D$-term requires a large difference between
$\mn$ and $\mnbar$ (of the same order of the slepton masses). If
some symmetry of the K\"{a}hler potential guarantees that
$\mn$ and $\mnbar$ are equal at the tree level, a large difference between
them can still be generated by radiative corrections, especially if
\ux is broken by the same radiative corrections at some much lower energy.
We consider such a model in the next section.


\section*{IV. Large D-term corrections from radiative breaking of
$\mbox{U(1)}_{{\bf X}}$}

If the scalar masses are universal  at the Planck scale because of some
symmetry of the K\"{a}hler potential,
the difference between $\mn$
and $\mnbar$ can still be generated by radiative corrections
below the Planck scale if $N$ and $\Nbar$ couple to other fields differently.
An interesting case is that the \ux is also broken by the same radiative
corrections which modify $\mn$ and $\mnbar$,
i.e., $N$ and $\Nbar$ fields get VEVs when $\ms = {1 \over 2} (\mn
+ \mnbar)$ is renormalized to negative. In this case, $\md ={1\over 2}(
\mn - \mnbar) \simeq \mn$ which is
presumably comparable to the masses of the squarks
and sleptons, then the $D$-term correction the the sparticle spectrum
can be quite large. In what follows we consider a simple model
which will demonstrate this case.

We assume, for simplicity, $M_{10}=M_G$, and
beneath $M_G$, the particle contents are the usual ones
in the MSSM with  3 right-handed neutrinos, the additional \ux gauge field,
an $N$ and an $\Nbar$ fields discussed above which break the \ux when
they get nonzero VEVs, and 3 gauge singlets $S_k$, $k=1, 2, 3$. The
$N$ and $\Nbar$ belong to the $16$ and $\overline{16}$ representations of
$SO(10)$ at the GUT scale with all other components get superheavy masses
and decouple below the GUT scale. This can be achieved by a $45$ Higgs
with VEVs in the hypercharge direction (see Appendix A). The two low
energy Higgs doublets $H_1$ and  $H_2$
are assumed to belong to the $10$ representations
of $SO(10)$ and their X charges are -2 and 2 respectively. The X charges
of all chiral fields are shown in Table 1.
Note that we only add the \sm
singlets to the MSSM so that the successful prediction of $\sin^2 \theta_W$
in the SGUTs is retained.

\begin{table}[htbp]
$$
\begin{array}{||c||c|c|c|c|c|c|c|c|c|c|c||} \hline
\mbox{field:} & q_L & {u_R}^c & {d_R}^c & l_L & {e_R}^c & {\nr}^c & H_1 & H_2 &
N & \Nbar & S \\ \hline
X & -1 & -1 & 3 & 3 & -1 & -5 & -2 & 2 & -5 & 5 & 0 \\ \hline
\end{array}
$$
\caption{The \ux charges of different fields}
\end{table}

We consider a  superpotential given by
\ba
W & = & \mbox{\boldmath $Q \lambda_U U^c$} H_2 + \mbox{\boldmath
     $Q \lambda_D D^c$}
      H_1 + \mbox{\boldmath $L \lambda_E E^c$} H_1 + \mbox{\boldmath
      $L \lambda_\nu \nr^c$} H_2 \samepage \\ & & + \mu H_1 H_2
      + \sum_{k=1}^3 \lambda_k {\nr}^c_k S_k \Nbar .
      \eqno{(4.1)}
\ea
Other possible interactions, such as $N S_k \Nbar$, $m S_k^2$
and $S_k^3$, could vanish
either because $S_k$'s are embedded in some non-trivial representations of
$SO(10)$, or because of some discrete symmetry. (For example, a parity
whose lepton fields change sign and $S_k$ and $\Nbar$ are multiplied by $i$.)
The scalar potential involving $N$ and $\Nbar$ fields is given by
\footnote{We use $S$ and $N$ to represent both the superfields and their
scalar components. It should be clear which one they represent.}
\ba
V & = & {1 \over 2} g_X^2 (X_N \abs{N}^2 + X_{\Nbar} |{\Nbar}^2| +
    \sum_i X_i \abs{\phi_i}^2)^2 +
    \sum_{k=1}^3 \abs{\lambda_k {{\snr}{}^c}_k  \Nbar}^2
    \samepage\\
    & &  + \sum_{k=1}^3 \abs{\lambda_k S_k \Nbar}^2 + \mn \abs{N}^2 + \mnbar
    |{\Nbar}^2|+ \sum_{k=1}^3 A_k \lambda_k {{\snr}{}^c}_k S_k \Nbar
    \samepage\\
   & = & {1 \over 2} g_X^2 (X_N \Delta +  \sum_i X_i \abs{\phi_i}^2)^2
    + \ms \Sigma + \md \Delta \samepage\\  & &
    + \sum_{k=1}^3 \abs{\lambda_k {{\snr}{}^c}_k \Nbar}^2
     + \sum_{k=1}^3 \abs{\lambda_k S_k \Nbar}^2
    + \sum_{k=1}^3 A_k \lambda_k {{\snr}{}^c}_k S_k \Nbar , \eqno{(4.2)}
\ea
where $\Sigma$, $\Delta$, $\ms$, and $\md$ are defined as before.
When $\ms$ is driven negative by  the Yukawa interactions
$\lambda_k {\nr^c}_k S_k \Nbar$ at some intermediate mass scale $M_I$,
($\lambda_k$'s are assumed to be ${\cal O}(1)$,)
$N$ and $\Nbar$ fields will get nonzero VEVs and break the \ux .
The difference of the squares of their VEVs $\Delta$ is given by
$\Delta = -{\md \over X_N^2 g_X^2}$ by minimizing $V$ with respect to $\Delta$,
and the sum $\Sigma$ is fixed by the one-loop correction
$$
\Delta V = {1 \over 64 \pi^2} Str M^4 [\ln{M^2 \over \mu^2} - {3\over 2}]
\eqno{(4.3)}
$$
to the scalar potential \cite{cw},
 $\Sigma \sim M_I^2$ where $M_I$ is the scale at
which $\ms(M_I) = \mn(M_I) + \mnbar(M_I) = 0$ \cite{hk}.
Fig. 1 shows the
evolutions of the soft breaking masses of $N$, $\Nbar$, $S_k$, and
${{\snr}{}^c}_k$
fields.
For simplicity, we have assumed that the soft SUSY breaking parameters
are universal at $M_G$ and
the parameters are chosen to be $\lambda_{t0}=
\lambda_{\nu_{\tau}0}=1.5,\, \lambda_{b0, \tau 0} \ll 1,\,
\lambda_{k0} =1,\, k=1,2,3$,
and the universal soft breaking trilinear couplings $A_0= 3 m_0$.
The $m_{{\snr}{}_{3}}^2$ is also driven negative at low energies
because of the large $\lambda_{\nu_3}$ coupling. However, the terms
$\sum_{k=1}^3 \abs{\lambda_k {{\snr}{}^c}_k  \Nbar}^2$ in the scalar
potential $V$
(equation(4.2)) prevent both $\Nbar$ and ${\snr}{}^c$ get non-zero VEVs.
After \ux is broken,
the mass square of ${\snr}{}^c_{3}$ gets a large positive contribution
from the $\Nbar$ VEV and $\vev{{\snr}{}^c_3}$ remains zero.

The present bounds on the mass of the \ux gauge boson $Z_\chi$ are
$M_{Z_\chi} > 320$ GeV (direct) and $> 670$ GeV (indirect) \cite{zprime}.
The primordial
nucleosynthesis may put a more stringent limit on $M_{Z_\chi}$, taking
$N_\nu < 3.5$, $M_{Z_\chi}$ has to be greater than ${\cal O}$(TeV) \cite{gv}
because of the extra massless states present in our model. Cosmological
constraints also put an upper limit on $M_I$. The flaton
(a linear combination of
$N$ and $\Nbar$ which corresponds to the quasi-flat direction) decays into
light particles through the heavy intermediate states of ${\cal O}(M_I)$
after the
phase transition of \ux breaking. The decay rate must be fast enough in
order not to affect the primordial nucleosynthsis or over-dilute the
baryon asymmetry. This gives an upper bound on $M_I$ \cite{u1pt}.
With these considerations, we will take $M_I$ to be in the range of
$10^3$GeV to $10^7$GeV.

Compared with MSSM, the scalar masses contain two
extra contributions: the \ux gaugino contribution and the \ux $D$-term.
For the first two generations where the Yukawa couplings are negligible,
the scalar masses are given by
\ba
m_i^2 & = & m_0^2 + \sum_{a=1}^3 f_{ai}M_0^2 + f_{Xi} M_0^2 \samepage
\\  \samepage
& & + (T_{3i} -
Q_i \sin^2 \theta_W)M_Z^2 \cos 2\beta - {X_i\over X_N}\md , \eqno{(4.4)}
\ea
where $m_0$ and $M_0$ are the scalar mass and gaugino mass at $M_G$
respectively, $f_{ai},\, a=1, 2, 3$ are the same as before and $f_{Xi}$ is
given by
$$
f_{Xi}  =  {2 \over b_X} {X_i^2 \over 40}[{\alpha_X^2(M_I) \over
\alpha_X^2(M_G)}-1] , \eqno{(4.5)}
$$
In this simple model, $\md$ can also be expressed in
terms of $m_0$ and $M_0$,
$$
\md = \mn = m_0^2 + f_{XN} M_0^2 , \eqno{(4.6)}
$$
then we have
\ba
m_i^2 & = &  (1-{X_i \over X_N})m_0^2 + \sum_{a=1}^3 f_{ai}M_0^2
+ (f_{Xi}-{X_i \over X_N}f_{XN}) M_0^2
\samepage\\ & &  + (T_{3i} - Q_i \sin^2 \theta_W)
M_Z^2 \cos 2\beta . \eqno{(4.7)}
\ea
The corrections $-{X_i \over X_N}m_0^2+ (f_{Xi}-{X_i \over X_N}f_{XN}) M_0^2$
to
the masses of squarks and sleptons compared to the MSSM can be as large as
60\% for $X_i=3$ in the limit $m_0 \gg M_0$. Fig. 2 shows the
comparison of the scalar spectra with  and without the \ux $D$-term corrections
for a set of $m_0$ and $M_0$. We see that the corrections are more significant
for the sleptons than for the squarks because of the smaller gaugino mass
contributions to the sleptons than to the squarks.
Now the slepton mass relation (3.1) is modified to be
\ba
m_E^2 - m_{E^c}^2& = &(C_2 - {3\over 4}C_1)M_0^2 + 8 C_X M_0^2 + {4\over 5}\md
\\ & & + (-{1\over 2} + 2 \sin^2
\theta_W)M_Z^2 \cos 2\beta, \eqno{(4.8)}
\ea
where $f_{Xi}= X_i^2 C_X$.
In a more general $SO(10)$ theory there is no simple relation between
$\md$ and $m_0^2$ and $\md$ has to be treated as a parameter.

Before going to the next section, we have three comments on this model.

(1) $S$ term contributions: When \ux is broken at intermediate energy $M_I$,
 $S(M_I)$ is also shifted
by $\delta m_{H_2}^2 -\delta m_{H_1}^2 = {4\over 5} \md$.
then the equations (2.13),
(2.17) and (2.18) are not valid. Therefore, if (2.15a),
(2.15b) hold but (2.18) does not, it may be a hint of an \ux breaking at
intermediate energy scale and providing a shift of the $S$ term.

(2) Neutrino masses: In our simplest model, there are three heavy Dirac
neutrinos and three massless neutrinos because of the three singlet
states we introduced \cite{heavynu,wr}.
We can see them from the mass terms of the neutrinos (for simplicity, we only
consider one family here and drop the family indices)
$$
m_D \nu_L \nr^c + M_D S \nr^c ,\eqno{(4.9)}
$$
where $m_D= \lambda_\nu \vev{H_2^0} \sim {\cal O}(m_{u,c,t})$, and $M_D=
\lambda
\langle{\Nbar}\rangle \sim {\cal O}(M_I)$.
One linear combination of $\nu_L$ and $S$,
$\nu_L \sin \theta + S \cos \theta$, where $\tan \theta = {m_D \over M_D}$,
is married with $\nr$ and gets a large mass $\sqrt{m_D^2+M_D^2} \sim
{\cal O}(M_I)$, which is consistent with experimental constraints \cite{wr},
and the other combination $\nu_L \cos \theta - S \sin \theta$ is left
massless. However, it is possible to give the three light neutrinos small
majorana masses which are favored to solve the solar neutrino problem
by just adding some extra interactions to the superpotential of the model.
For example, if we add to the superpotential the non-renormalizable
interaction ${1 \over M_G} S^2 N \Nbar$ which gives a small majorana mass term
$m_S S^2 = {1 \over M_G} \vev{N} \langle{\Nbar}\rangle S^2$
to $S$, then the mass matrix of the fields $\nu_L$, $\nr^c$, and $S$
becomes
$$
{\cal M} = \left( \begin{array}{ccc}
                   0 & m_D & 0 \\
                   m_D & 0 & M_D \\
                   0 & M_D & m_S
                  \end{array}  \right). \eqno{(4.10)}
$$
The product of the three mass eigenvalues is given by $\det {\cal M}
= - m_D^2 m_S$, and the two larger mass are approximately equal to $M_D$,
so the mass of the light neutrino is approximate
$$
m_{light} \simeq {m_D^2 m_S \over M_D^2} \sim {m_D^2 M_I^2 \over M_D^2 M_G}
\sim {m_D^2 \over M_G} \eqno{(4.11)}
$$
which is similar to that generated by the usual see-saw mechanism.

(3) $b$-$\tau$ Yukawa unification: Because the \ux is broken at low energy,
there are extra interactions surviving at low energies compared with
the MSSM. Especially the $\tau$-neutrino Yukawa coupling $\lambda_{\nu_\tau}$
which should be about the same as $\lambda_t$ at the GUT scale enters
the RG equations of many parameters. The RG equation for the $b$-$\tau$ mass
ratio $R$ is modified to be
$$
{dR \over dt} = {R \over 16\pi^2}(-{16\over 3}g_3^2 + {4\over 3}g_1^2
+\lambda_t^2 - \lambda_{\nu_\tau}^2 + 3 \lambda_b^2 - 3\lambda_{\tau}^2).
\eqno{(4.12)}
$$
In the small $\tan \beta$ case where $\lambda_b$ and $\lambda_\tau$ can
be neglected, the unification of $b$ and $\tau$ Yukawa couplings in SGUT
requires a large top Yukawa coupling to compensate the contribution
from the $SU(3)$ gauge coupling. In our model the contribution
of $\lambda_t$ is largely cancelled out by $\lambda_{\nu_\tau}$, making
it difficult to achieve the $b$-$\tau$ unification for the top Yukawa
coupling staying in the perturbative regime at the GUT scale. However,
since the $b$- and $\tau$- Yukawa couplings are small, they do not necessarily
come from a single renormalizable interaction of the form
\mbox{$16_3\, 10 \, 16_3\,$}
in $SO(10)$ and therefore their unification is not mandatory. In the
large $\tan \beta$ case where $\lambda_b$ and $\lambda_\tau$ are comparable
to $\lambda_t$ (which we will discuss in the next section),
the terms $3\lambda_b^2 - 3 \lambda_\tau^2$ in the RG
equation for $R$ also contribute and make up the negative contribution
from $\lambda_{\nu_\tau}$ ($\lambda_b > \lambda_\tau$ below the GUT scale).
In addition, the couplings between $b$ and $H_2$ through
the bottom squark-gluino loops and top squark-chargino loops \cite{copw,hrs}
could also
give a significant contribution to $R$ if $\tan \beta$ is large. Therefore,
the $b$-$\tau$ unification is possible in this case.

\section*{V. Fine-tuning problem in the Yukawa unification scenario}

Recently, the large $\tan \beta$ scenario in which the tau lepton and
the bottom and top quark Yukawa couplings unify at the grand unification
scale has drawn considerable interest \cite{hrs,yuni,rewsb}.
This happens in an $SO(10)$ GUT
if the two light Higgs
doublets lie predominantly in a single 10 representation of the gauge
group $SO(10)$ and the $t$, $b$, and $\tau$ masses originate in the
renormalizable Yukawa interactions of the form $16_3\, 10\, 16_3$.
In this case, the top quark mass can also be predicted and it was
predicted to be heavy \cite{hrs}. In fact, such a heavy top quark
is favored by the recent CDF results, $m_t = 174^{+10 +13}_{-10 -12}$ GeV
\cite{CDF}.
The problem with this scenario
is that radiative electroweak symmetry breaking is hard to achieve
although significant progress has already been made \cite{rewsb,hrs2,mn,op}.
The masses of the up- and down-type Higgs are the same at $M_{10}$ because they
lie in the same representation and run almost in parallel
because of the boundary condition $\lambda_t(M_{10}) = \lambda_b(M_{10})$.
Usually one relies on heavy gauginos to amplify the small hypercharge-induced
difference in the running of $m_{H_1}^2$ and $m_{H_2}^2$. However, all
these attemps require severe fine tuning of the parameters which we will
explain below.

The relevant part of the Higgs potential is given by
\ba
\mu_1^2 {|H_1^0|}^2 + \mu_2^2 {|H_2^0|}^2 + B\mu (H_1^0 H_2^0 +
\mbox{h.c.}) + {1\over 8}(g^2+ g'^2)({|H_1^0|}^2 - {|H_2^0|}^2)^2 .
\eqno{(5.1)}
\ea
Minimizing  the Higgs potential we obtain the following conditions,
\ba
{\mu_1^2 - \tan^2 \beta \mu_2^2 \over \tan^2 \beta - 1}  =
{M_Z^2 \over 2},  \eqno{(5.2)} \\
{-\mu B \over \mu_1^2 + \mu_2^2}  =  {1 \over 2} \sin 2\beta . \eqno{(5.3)}
\ea
In the case of $\lambda_t(M_{10}) = \lambda_b(M_{10})$
, $\tan \beta \simeq {m_t \over
m_b} \sim {\cal O}(50) \gg 1$. We see that $\mu_2^2 \simeq
-{M_Z^2\over 2}$ for $\mu_1^2$ not too large, then
$$
m_A^2=\mu_1^2 + \mu_2^2 \simeq (\mu_1^2 - \mu_2^2) - M_Z^2 < \mu_1^2 -
\mu_2^2 \equiv \epsilon_c m_S^2 , \eqno{(5.4)}
$$
where $m_A$ is the CP-odd scalar mass ,
$m_S^2$ is the typical supersymmetric particle mass scale,
$m_S \sim \max (m_0, M_0)$, and $\epsilon_c$ represents the custodial symmetry
breaking effects. Equation(5.4) tells us that both $m_A^2$ and $M_Z^2$ are
smaller than $\epsilon_c m_S^2$, so there is an ${\cal O}(\epsilon_c)$
fine-tuning
of the Z mass. In addition, writing $m_A^2 = \epsilon m_S^2$, $\epsilon <
\epsilon_c \ll 1$, we have
$$
{-\mu B \over \epsilon m_S^2} = {1 \over 2}\sin 2\beta \simeq {1 \over
\tan \beta} \\ \Rightarrow \\-\mu B \simeq {\epsilon \over \tan \beta} m_S^2 .
\eqno{(5.5)}
$$
While $\mu$ is typically of the order $m_S$ in order to satisfy
$\mu_2^2 = m_{H_2}^2 +\mu^2 \simeq -{M_Z^2 \over 2}$. The $B$ parameter
which receives contributions from the gaugino masses and the
soft SUSY-breaking trilinear scalar coupling $A$ and therefore is also
naturally of the order $m_S$ has to be fine-tuned to ${\cal O}({\epsilon \over
\tan \beta} m_S)$. The fine-tuning is at least one part in $10^3$
and is much worse than the naive
expectation ${1 \over \tan \beta}$.

The \ux $D$-term which gives the opposite contributions to $\mhd$ and $\mhu$
provides the desired ingredient to solve  this problem \cite{hrs2,hem}.
One can either simply have $\mn \neq \mnbar$ at tree level \cite{hrs2}
 or have the
difference $\md$ generated by radiative corrections as described in the
last section. However, the simple
model discussed in the previous section gives a positive contribution to
$\mhu$ and a negative contribution to $\mhd$ which is incompatible with
the fact that $\mu_1^2 > \mu_2^2$. We thus modify the model so that it
has interactions ${\lambda_k}' {\nr^c}_k {S_k}' N,\, k=1, 2, 3$, instead of
$\lambda_k {\nr^c}_k S_k \Nbar$. The ${S_k}'$'s are still \sm singlets,
but carry \ux charge $+10$ (they may belong to the $\overline{126}$ of SO(10)).
We also have to add $\overline{{S_k}}' (X=-10)$ to the model in order to
cancel the anomaly and we assume that they only have the \ux gauge
interaction. Then, the $\mn$, instead of $\mnbar$, is driven negative by
the Yukawa interactions. The $\md = {1 \over 2}(\mn - \mnbar)$ becomes
negative in this case and therefore it gives the correct-sign $D$-term
contributions to $\mhd$ and $\mhu$. Let $\delta m_H^2$ be the difference
between $\mhd$ and $\mhu$ generated by the renormalization group from
$M_{GUT}$ to $m_S$ without the $D$-term correction.
$$
\delta m_H^2 = \mhd -\mhu = \epsilon_c m_S^2 \ll m_S^2. \eqno{(5.6)}
$$
The parameter $m_A^2 = \mu_1^2 +\mu_2^2$ is now given by
$$
m_A^2 = \mu_1^2 +\mu_2^2 = (\mu_1^2 - \mu_2^2) +2 \mu_2^2
\simeq D + \delta m_H^2 - M_Z^2, \eqno{(5.7)}
$$
where $D=(-{2 \over 5}\md) - (+{2 \over 5}\md) = -{4 \over 5}\md
\sim m_S^2$. For $m_S$ larger than $M_Z$,
$m_A^2$ is naturally of the order $m_S^2$ and the problem of a light
$m_A^2$ can be avoided. The fine-tuning problem of $\mu B$ is also relieved
though not totally eliminated as we can see from equation(5.5) that a fine
tune of ${1\over \tan \beta} \sim {\cal O}({1\over 50})$ is still required.
However,
it should be generic since a large pure number $\tan \beta$ has to be
generated.

\section*{VI. Conclusions}

It is well known that quark and lepton mass and mixing angle relations may
provide evidence for grand unification. Although squarks and sleptons have yet
to be discovered, mass relations amongst scalars provide a much more reliable
test of unification than do the relations involving fermion masses. This is
because chiral and gauge symmetry breaking effects mask the grand unified
symmetry relations for the fermions, but are not present for the scalars.
In this paper we have derived several scalar mass relations which follow
directly from the grand unified symmetry, and we have studied the reliability
of such relations as a probe of supersymmetric unification.

The small size of flavor-changing processes suggests that in models with
weak-scale
supersymmetry the squarks of a given charge should be approximately
degenerate. This has led to the speculation that squarks and sleptons of
different charge might also be degenerate. Although only a speculation, such a
boundary condition of universal scalar masses has become a ubiquitous feature
of supersymmetric models and is incorporated in the minimal supersymmeric
standard model. Since there are five types of quark and leptons, the quark and
lepton weak doublets $Q$ and $L$ and the weak singlets $U^c$,$D^c$ and $E^c$,
such a boundary condition leads to four relations between the scalar masses.
However, the origin of these relations is more a matter of simplicity than
of any underlying fundamental principle.

In this paper we have derived mass relations, between scalars of a given
generation, which result from the most general possible boundary condition that
respects a grand unified symmetry.
With $SU(5)$ unification, the five types of
quarks and leptons are unified into two irreducible representations
($Q,\, U^c,\, E^c$) and ($L,\, D^c$),
leading to the expectation of three mass relations, which
are given in equation (2.14). However, these 3 relations involve a
quantity $T$, which depends on the mass splitting of the Higgs scalars at the
unification mass. It is likely that this mass splitting is small enough that
the relations (2.14) with $T=0$ will result. However, if the mass splitting is
very large there are only 2 mass relations between the scalar mass parameters
of each of the light generations. These relations are given by eliminating $T$,
and are given in equations (2.15). We believe that these relations must be
correct in any grand unified theory which incorporates the usual $SU(5)$ group.
If these relations are found to be incorrect, then it is unlikely that grand
unification is correct. Although extra particles and interactions could be
added to a grand unified theory to invalidate these mass relations, such
particles and interactions will lead to extra renormalizations of the weak
mixing angle, upsetting the outstanding agreement between the theoretical
prediction and the experimental value.

Even if the parameter $T$ is large, a
third mass relation can be derived because $T$ can be
evaluated by measuring the Higgs boson and third generation scalar masses. This
mass relation is given in equation (2.18).

If the quark and leptons are further unified, so that all 5 species of a
generation are unified in a single representation, as occurs in $SO(10)$
theories, a fourth mass relation is
to be expected. This is written, ignoring $T$, in equation (3.1), as a
relation between the masses of the two charged sleptons. This mass
relation is likely to be the first which is subject to precise experimental
test. If it were verified it would provide striking support for SO(10)
unification. However, unlike the two mass relations mentioned above, it is not
a necessary consequence of
$SO(10)$ unification. We have shown in this paper that
it is possible to have large corrections to this mass relation from $U(1)_X$
$D^2$ interactions, either at tree level or by radiative corrections.

\section*{Acknowledgments}

H.-C. Cheng would like to thank Hitoshi Murayama for many useful discussions.
This work was supported in part by the Director, Office of Energy Research,
Office of High Energy and Nuclear Physics, Division of High Energy Physics
of the U.S. Department of Energy under Contract No. DE-AC03-76SF00098 and
in part by the National Science Foundation under Grant No. PHY-90-21139.


\appendix

\section*{Appendix A}

In this appendix we show that it is possible to give superheavy masses to
all components but the \sm singlet of a multiplet of $SO(10)$. This can be
achieved by a 45 of $SO(10)$, $A_Y$, with VEV in the hypercharge direction.
The interaction,
$$
\overline{C} \, A_Y \, C ,\eqno{(A.1)}
$$
where $C = 16$ or $126$, will give superheavy masses to the components
of $C$ and $\overline{C}$ which have non-zero hypercharges, and leave the
\sm  singlets ($N$, $\Nbar$, ${S_k}'$, $\overline{S_k}'$) massless.
Those singlets
survive below the GUT scale and serve to break the \ux at low energy.

To generate a 45 VEV in the hypercharge direction, we start with the
following $SO(10)$ invariant superpotential, (we denote 54 of $SO(10)$
by $S$, 45 by $A$, and the singlet by $\chi$)
$$
W_1 = m_1 A_1^2 + m_2 S^2 + m_3 \chi^2 + \lambda_1 A_1^2 S + g_1 A_1^2 \chi .
\eqno{(A.2)}
$$
The equations for a supersymmetric minimum are
\ba
0 & = & F_A = 2 (m_1 + \lambda_1 S + g_1 \chi) A_1 \\
0 & = & F_S = 2 m_2 S + \lambda_1 (A_1^2 - {1 \over 10} Tr A_1^2) \\
0 & = & F_\chi = 2 m_3 \chi + g_1 Tr A_1^2 . \eqno{(A.3)}
\ea
Choosing
\ba
\vev{S} & = & \left( \begin{array}{cc} 1 & 0 \\ 0 & 1 \end{array}
                   \right) \otimes \mbox{diag} (s, s, s, -{3\over 2}s,
                   -{3\over 2}s),\\
\vev{A} & = & \left( \begin{array}{cc} 0 & 1 \\ -1 & 0 \end{array}
                   \right) \otimes \mbox{diag} (a, a, a, b, b),\\
\mbox{and}\;\;\;\;\;  \vev{\chi} & = & c, \eqno{(A.4)}
\ea
the above equations become
\ba
(m_1 + \lambda_1 s + g_1 c) a & = & 0 \\
(m_1 - {3\over 2} \lambda_1 s + g_1 c) b & = & 0 \\
2 m_2 s + {2\over 5} \lambda_1 (b^2-a^2) & = & 0 \\
2 m_3 c + g_1 (6 a^2 + 4 b^2) & =& 0 . \eqno{(A.5)}
\ea
We are interested in the solution $s=0,\, a=b=({m_3\over 5g_1}c)^{1\over 2}
\neq
0,\, c=-{m_1 \over g_1}$, in which $\vev{A_1}$ is in the $SU(5)$ singlet
($X$-charge) direction. We have obtained the breaking pattern $SO(10)
\rightarrow SU(5) \times U(1)_X$. We next add to the theory in such a way
that $SU(5)$ breaks to the gauge groups of the standard model, and the only
light states beneath the GUT scale are those  of the MSSM with some standard
model singlets.
We add the following terms to the superpotential.
\ba
W_2 & = & g_2 \chi' A_1 A_Y + \lambda_2 S' A_Y^2 + m_4 S' S_Y
       + m_5 S_Y^2 + \lambda_3 S_Y^3 \\
    &  & + \lambda_4 A A_1 A_Y + m_6 A^2 . \eqno{(A.6)}
\ea
We assume that ${\chi}'$ and $S'$ have no VEVs so that the minimization of
$W_1$ is not affected, and $S_Y$ and $A_Y$ have
non-zero VEVs of the following forms,
\ba
\vev{S_Y} & = & \left( \begin{array}{cc} 1 & 0 \\ 0 & 1 \end{array}
                   \right) \otimes \mbox{diag} (s_Y, s_Y, s_Y, -{3\over 2}s_Y,
                   -{3\over 2}s_Y),\\
\vev{A_Y} & = & \left( \begin{array}{cc} 0 & 1 \\ -1 & 0 \end{array}
                   \right) \otimes \mbox{diag} (a_Y, a_Y, a_Y, b_Y, b_Y).
\eqno{(A.7)}
\ea
The $s_Y$ is determined by the equation
$$
0 = F_{S_Y} = m_4 S' + 2 m_5 S_Y + 3 \lambda_3 (S_Y^2 - {1 \over 10}
Tr S_Y^2). \eqno{(A.8)}
$$
We obtain $ s_Y = {4 m_5 \over 3 \lambda_3}$.
The $F_{\chi'}$ equation,
$$
F_{\chi'}=A_1 A_Y =0, \eqno{(A.9)}
$$
force $\vev{A_Y}$ to be in the direction orthogonal to $\vev{A_1}$ and the
$F_{S'}$ equation,
$$
F_{S'} = \lambda_2 (A_Y^2 - {1\over 10} Tr A_Y^2) + m_4 S_Y = 0, \eqno{(A.10)}
$$
force $\vev{A_Y}$ to be in the hypercharge direction. We have
$$
b_Y = -{3\over 2} a_Y \;\;\; \mbox{and} \
;\;\; a_Y^2 = {2 m_4 s_Y \over \lambda_2}
\eqno{(A.11)}
$$
from these equations. Finally, we need the trilinear interaction
$\lambda_4 A A_1 A_Y$  to make sure that there are no
extra massless states which are not eaten by the gauge bosons present to
destroy the successful $\sin^2 \theta_W$ prediction. We have checked
the mass matrices of these fields and indeed there are only 32 massless
modes which are needed for the symmetry breaking $SO(10) \rightarrow
SU(3)_c \times SU(2)_L \times U(1)_Y \times U(1)_X$.
Now we have successfully constructed a superpotential
which generates 45 VEV in the hypercharge direction. The
\sm  singlet- non-singlet splitting required in our model could be obtained
from the interaction (A.1) consequently.

\newpage

\section*{Figure Captions}

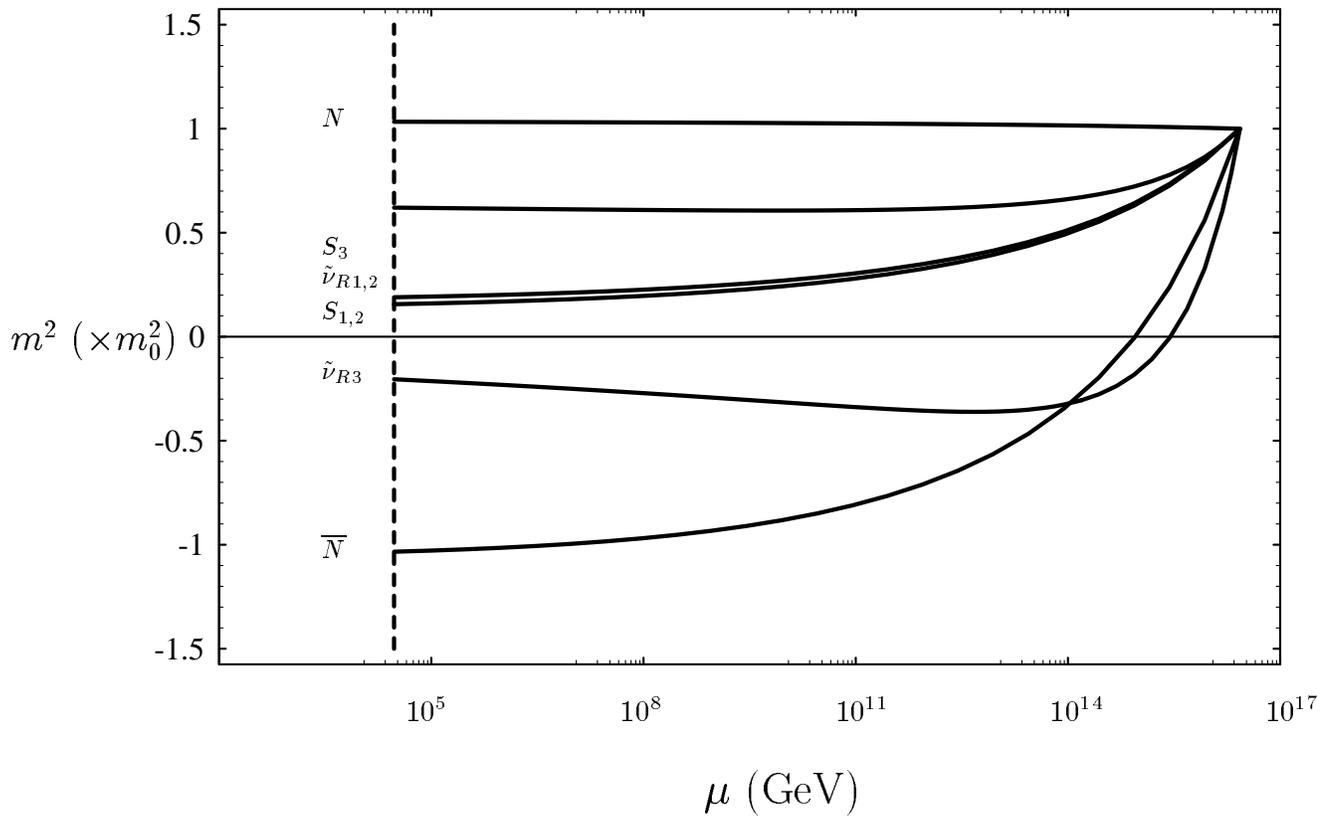
\noindent Fig. 1.  The evolutions of the soft breaking masses of
$N$, $\Nbar$, $S_k$, and ${{\snr}{}^c}_k$ fields from GUT scale
($2.7 \times 10^{16}$ GeV) to \ux
breaking scale (30 TeV).
An universal soft breaking mass $m_0$ is assumed at GUT scale and
the parameters are chosen to be $\lambda_{t0}=
\lambda_{\nu_{\tau}0}=1.5, \lambda_{b0, \tau 0} \ll 1, \lambda_{k0} =1,
k=1,2,3$
, and the universal soft breaking trilinear couplings $A_0= 3 m_0$.

\vskip .3in
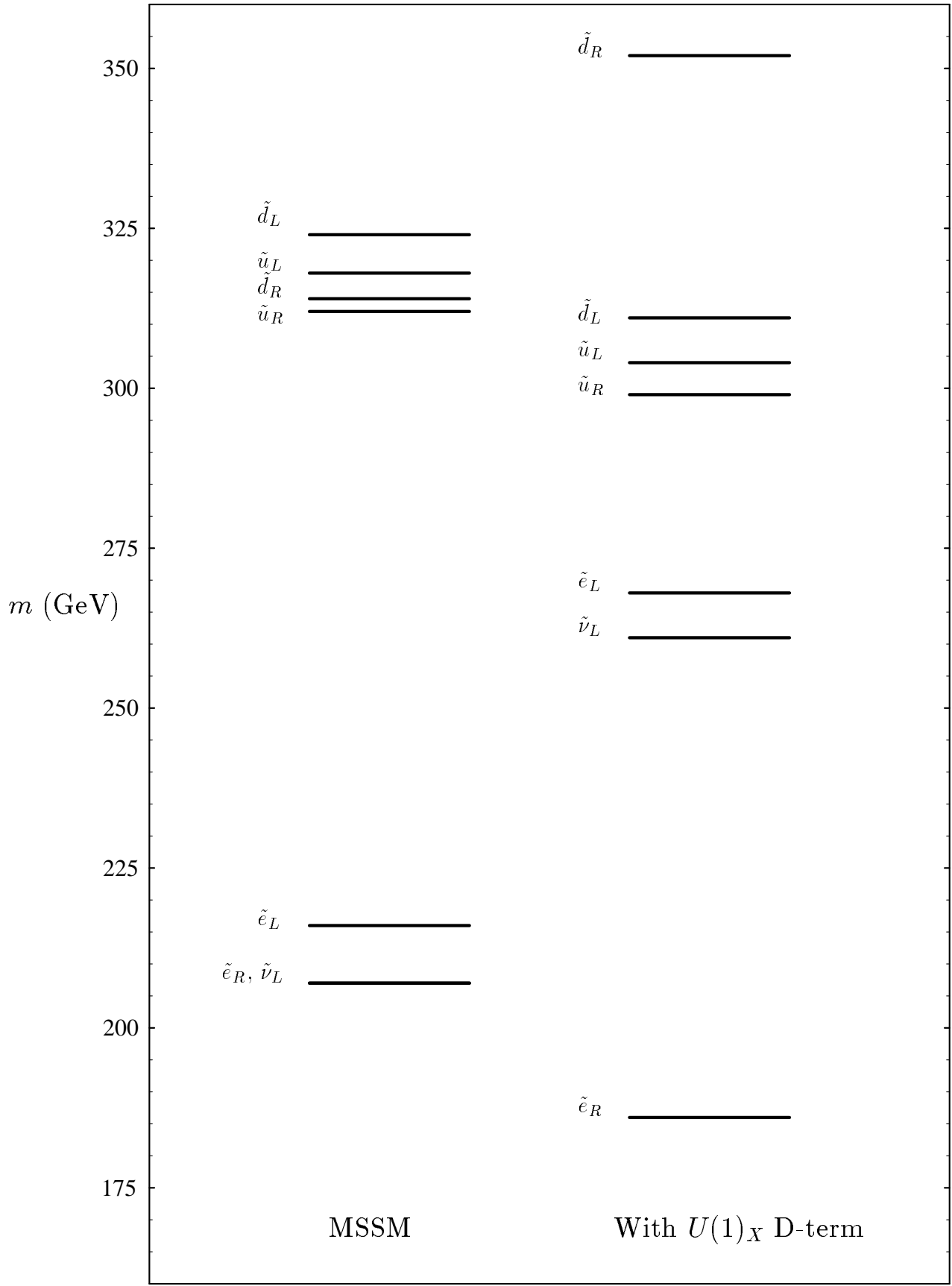
\noindent Fig. 2. Comparison of the scalar particle spectra with
and without the \ux $D$-term corrections
for a set of $m_0$, $M_0$ and $\tan \beta$.
\end{document}